\documentclass[a4paper,11pt]{article}
\usepackage{pos}
\usepackage{amsmath}
\usepackage{graphicx}
\usepackage{hyperref}
\usepackage{multicol}

\title{The $\Lambda(1405)$ from Lattice QCD: \\ Determining the Finite-volume Spectra}

\author[a]{John Bulava}
\author*[b,c]{B\'{a}rbara Cid-Mora}
\author[d]{Andrew D. Hanlon}
\author[e]{Ben H\"{o}rz}
\author[c,b]{Daniel Mohler}
\author[f]{Colin Morningstar}
\author[g]{Joseph Moscoso}
\author[g]{Amy Nicholson}
\author[h]{Fernando Romero-L\'{o}pez}
\author[f]{Sarah Skinner}
\author[i]{Andr\'{e} Walker-Loud}

\affiliation[a]{Fakult\"{a}t f\"{u}r Physik und Astronomie, Institut f\"{u}r Theoretische Physik II, Ruhr-Universit\"{a}t Bochum, 44780 Bochum, Germany}
\affiliation[b]{GSI Helmholtz Centre for Heavy Ion Research, Darmstadt, Germany}
\affiliation[c]{Institut f\"ur Kernphysik, Technische Universit\"at Darmstadt,
Schlossgartenstrasse 2, 64289 Darmstadt, Germany}
\affiliation[d]{Physics Department, Brookhaven National Laboratory, Upton, New York 11973, USA}
\affiliation[e]{Intel Deutschland GmbH, Dornacher Str. 1, 85622 Feldkirchen, Germany}
\affiliation[f]{Department of Physics, Carnegie Mellon University, Pittsburgh, Pennsylvania 15213, USA}
\affiliation[g]{Department of Physics and Astronomy, University of North Carolina, Chapel Hill, NC 27516-3255, USA}
\affiliation[h]{Center for Theoretical Physics, Massachusetts Institute of Technology, Cambridge, MA 02139, USA}
\affiliation[i]{Nuclear Science Division, Lawrence Berkeley National Laboratory, Berkeley, CA 94720, USA}

\emailAdd{b.cidmora@gsi.de}

\abstract{
     This work presents technical details of determining the finite-volume energy spectra for the scattering amplitude of the coupled-channel $\pi\Sigma - \bar{K}N$ from lattice QCD data. The importance of reliably extracting such spectra lies in the crucial dependence of the hadronic scattering amplitudes analysis on the energy spectrum when using L{\"u}scher's formalism. Results of the methods used are presented and the final finite-volume spectra are shown. The analysis of the scattering amplitude based on these results, exhibits a two-pole structure for the $\Lambda(1405)$, a virtual bound state below the $\pi\Sigma$ threshold and a resonance pole right below the  $\bar{K}N$ threshold.
}

\FullConference{%
The 40th International Symposium on Lattice Field Theory (Lattice 2023)\\
July 31st - August 4th, 2023\\
Fermi National Accelerator Laboratory\\
}

\begin{document}
\maketitle

\section{Introduction}

The $\Lambda(1405)$ baryon with strangeness $S=-1$, isospin $I=0$, negative parity and spin $J=1/2$ is recognized as a 4-star resonance by the Particle Data Group \cite{ParticleDataGroup:2022pth}, implying a somewhat well-known hadron. It was first predicted in Ref.~\cite{Dalitz:1959dn}, where the existence of a resonance in the $\pi^{-}\Sigma^{+}$ spectrum was suggested just below the $K^{-}p$ threshold. Historically, this hadron attracted attention due to its unexpectedly low mass compared to its nucleon counterpart $N(1535)$ when studied from a conventional quark model picture. Currently, the growing interest in investigating this baryon is rather based on the ideas proposed in Ref.~\cite{Fink:1989uk}, specifically the concept of two-pole structures. This concept allows the existence of an additional state, and cases like the $\Lambda(1405)$ can be dynamically generated by meson-baryon interactions \cite{Oller:2000fj, Jido:2003cb}. Although many lattice QCD studies have been conducted since, none of them have extracted the full spectrum in the $\pi\Sigma-\bar{K}N$ coupled-channel region in order to appropriately study the pole structure.

These aspects motivated this first study \cite{Bulava:2023rmn} determining the coupled-channel $\pi\Sigma-\bar{K}N$ scattering amplitudes from lattice QCD, thus attempting to get a better understanding of the pole structure and the positions of the resonance poles in the $\Lambda(1405)$ region. In order to achieve this, the L{\"u}scher formalism \cite{Luscher:1990ux,Luscher:1991cf} is employed, which relates discrete finite-volume energy spectra extracted from lattice QCD data to scattering amplitudes. Given the crucial dependence of the scattering amplitudes on the finite-volume spectra,  the latter must be extracted reliably from lattice QCD calculations. Therefore, the main focus of this report is to present essential specifics of the lattice QCD calculations, which include details about the generation of lattice data, type of operators used to construct correlation functions, analysis of correlation functions, and final extraction of finite-volume energy spectra. 

This brief introduction is followed by Section~\ref{section:Correlators} which summarizes all the details of the lattice QCD ensemble used, including the construction of correlation functions based on a diverse set of interpolating operators. This part also introduces the methodology used to diagonalize correlation matrices, its variations and example results. Section~\ref{section:FiniteVolumeSpectra} presents the fits performed to lattice data that led to the final results, meaning the final finite-volume energy spectra. Finally, Section~\ref{section:Conclusion} outlines the main results and conclusions from the methods used, as well as the importance for the analysis of scattering amplitudes.

\section{Spectrum determination}
\label{section:Correlators}

The determination of the finite-volume energy spectra from lattice QCD data starts with the generation of gauge configurations, followed by the evaluation of correlation functions using an appropriate operator basis, continued by diagonalization of these correlation functions, and finally fitting the data to extract the finite-volume energy spectra.

\subsection{Ensemble details}

The study is carried out using a single ensemble of QCD gauge configurations: D200, generated by the Coordinated Lattice Simulations (CLS) consortium \cite{Bruno:2014jqa}. The light quark masses used for the gauge fields generation are heavier-than-physical and degenerate $u-$ and $d-$quarks, and a lighter-than-physical $s-$quark. The resulting hadron masses, $m_{\pi}L$ and properties of the lattice are shown in Table~\ref{tab:LatticeInfo}. The gauge configurations of the D200 ensemble were generated using the tree-level improved L{\"u}scher-Weisz gauge action and a non-perturbatively $O(a)$-improved Wilson fermion action. Additionally, open temporal boundary conditions were employed in order to reduce autocorrelation, constraining the interpolating fields to be far from the boundaries, hence the maximal temporal separation in correlation functions of $t_{\rm max} = 25a$. 

\begin{table}[!h]
    \centering
    \begin{tabular}{cccccc}
    \hline\hline
    $a [\textup{fm}]$ & $(L/a)^{3} \times T/a$ &   $am_{\pi}$ & $am_{\rm K}$ & $m_{\pi}L$\\\hline 
    $0.0633(4)(6)$ & $64^{3} \times 128$ &  $0.06535(25)$  & $0.15602(16)$ & $4.181(16)$\\
    \hline\hline
    \end{tabular}
    \captionsetup{width=.95\textwidth}
    \caption{Properties of the D200 ensemble, where $a[\rm fm]$ and $(L^{3}/a) \times (T/a)$ are the lattice spacing and the lattice extent, respectively. The pion and kaon masses are $m_{\pi} \approx 200$ MeV and $m_{\rm K} \approx 487$ MeV.}
    \label{tab:LatticeInfo}
\end{table}

\subsection{Correlation functions}

The correlation functions are built using an appropriate set of interpolating operators that overlap with the states of interest, and include both single-hadron operators and multi-hadron operators \cite{Morningstar:2013bda}, as well as meson-baryon interpolators with different momentum combinations: $\Lambda(\Vec{\mathbf{d}}^{2})$, $\pi (\Vec{\mathbf{d}}_{1}^{2}) \Sigma (\Vec{\mathbf{d}}_{2}^{2})$ and $\bar{K} (\Vec{\mathbf{d}}_{1}^{2}) N (\Vec{\mathbf{d}}_{2}^{2})$, as shown in Table~\ref{tab:operators} (see Ref.~\cite{Bulava:2023gfx} for the complete list and for a comprehensive description of the naming scheme). The inclusion of a more diverse set of operators has the fundamental role of ultimately extracting the full finite-volume energy spectra below the lowest-three-particle threshold, namely the $\pi\pi\Lambda$ threshold.

\begin{table}[h!]
\centering
\begin{tabular}{c@{\hspace*{12mm}}l}
\hline
$\Lambda(\mathbf{d}^2)$ & Operators \\
\hline\hline
$G_{1 \rm g}(0)$ 
& $\Lambda[G_{1\rm g}(0)]_{0,1,3}$ \\
&  $\bar{K}[A_{2}(1)]_1\ N[G_1(1)]_0$\\
&  $\pi[A_{2}^{-}(1)]_1\ \Sigma[G_1(1)]_0$\\ \hline
\end{tabular}
\captionsetup{width=.95\textwidth}
\caption{Example of single- and multi-hadron operators used to construct correlation matrices. The irreducible representation ($\Lambda(\mathbf{d}^2)$) labels are related to the symmetry sector with a total momentum $\mathbf{d}^2$, and the subscript indicates a spatial identification number.}
\label{tab:operators}
\end{table}

\begin{figure}[ht!]
\centering
\includegraphics[width=.65\textwidth]{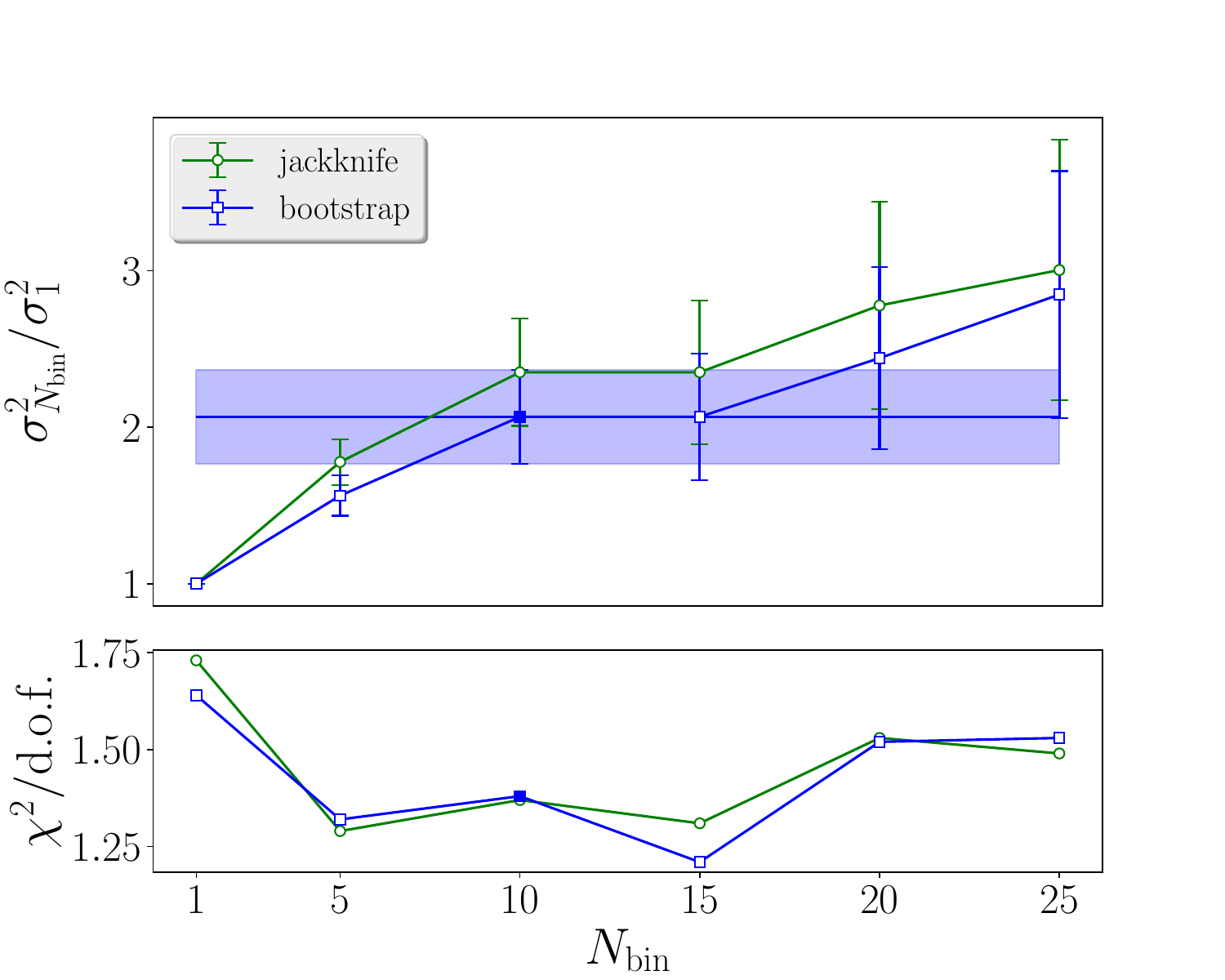}
\captionsetup{width=.9\textwidth}
\caption{(Top) Ratios of variances for fits to $m_{\pi}$ versus $N_{\rm bin}$ for jackknife and bootstrap resampling. (Bottom) Correlated-$\chi^{2}$ of two-exponential fit to $m_{\pi}$ versus $N_{\rm bin}$. In both panels, the final binning choice is illustrated as a blue solid square.}
\label{fig:rebin_analysis}
\end{figure}

These temporal correlation functions are evaluated using the stochastic Laplacian Heaviside method (sLapH) \cite{HadronSpectrum:2009krc, Morningstar:2011ka}. Once the correlation matrices are computed, autocorrelation of the data is studied with binning by computing the single hadron masses and their variance with different $N_{\rm bin}$ (see Fig.~\ref{fig:rebin_analysis} as an example of the pion mass). The final binning choice $N_{\rm bin}=10$ is made based on behavior of the variance and the corresponding correlated-$\chi^{2}$ for a certain value of $N_{\rm bin}$ of resampled data (jackknife or bootstrap), and these results are shown in Fig.~\ref{fig:rebin_analysis}. 

\subsection{Extraction of energy spectra}
\label{section:DiagonalizationCorrelators}

Now the correlation matrices must be diagonalized in order to extract stationary-state energies. This is achieved by solving the so-called Generalized Eigenvalue Problem (GEVP) (more details of this method can be found in Refs.~\cite{Michael:1982gb,Blossier:2009kd,Bulava:2022vpq}). The method diagonalizes correlation matrices as:
\begin{equation}
    C(t_{\rm d}) \Vec{v}_{n} (t_{0}, t_{\rm d}) = \lambda_{n} (t_{0}, t_{\rm d})\ C(t_{0})\ \Vec{v}_{n} (t_{0}, t_{\rm d}),
\end{equation}
where $t_{0}$ is the metric time, $t_{\rm d}$ is the diagonalization time, and $\lambda_{n}$ are the eigenvalues. This prescription connects the latter to an exponential of the form:
\begin{equation}
    \lambda_{n}(t, t_{0}) \propto \text{e}^{-E_{n}(t-t_{0})} \left(1 + \mathcal{O}({\rm e}^{-\Delta E_{n}(t-t_{0})})\right), 
\end{equation}
where $\Delta E_{n}$ is the distance to the closest energy level, and discussed in more details in Ref.~\cite{Blossier:2009kd}. 

\begin{figure}[hbt]
    \centering
    \includegraphics[width=.63\textwidth]{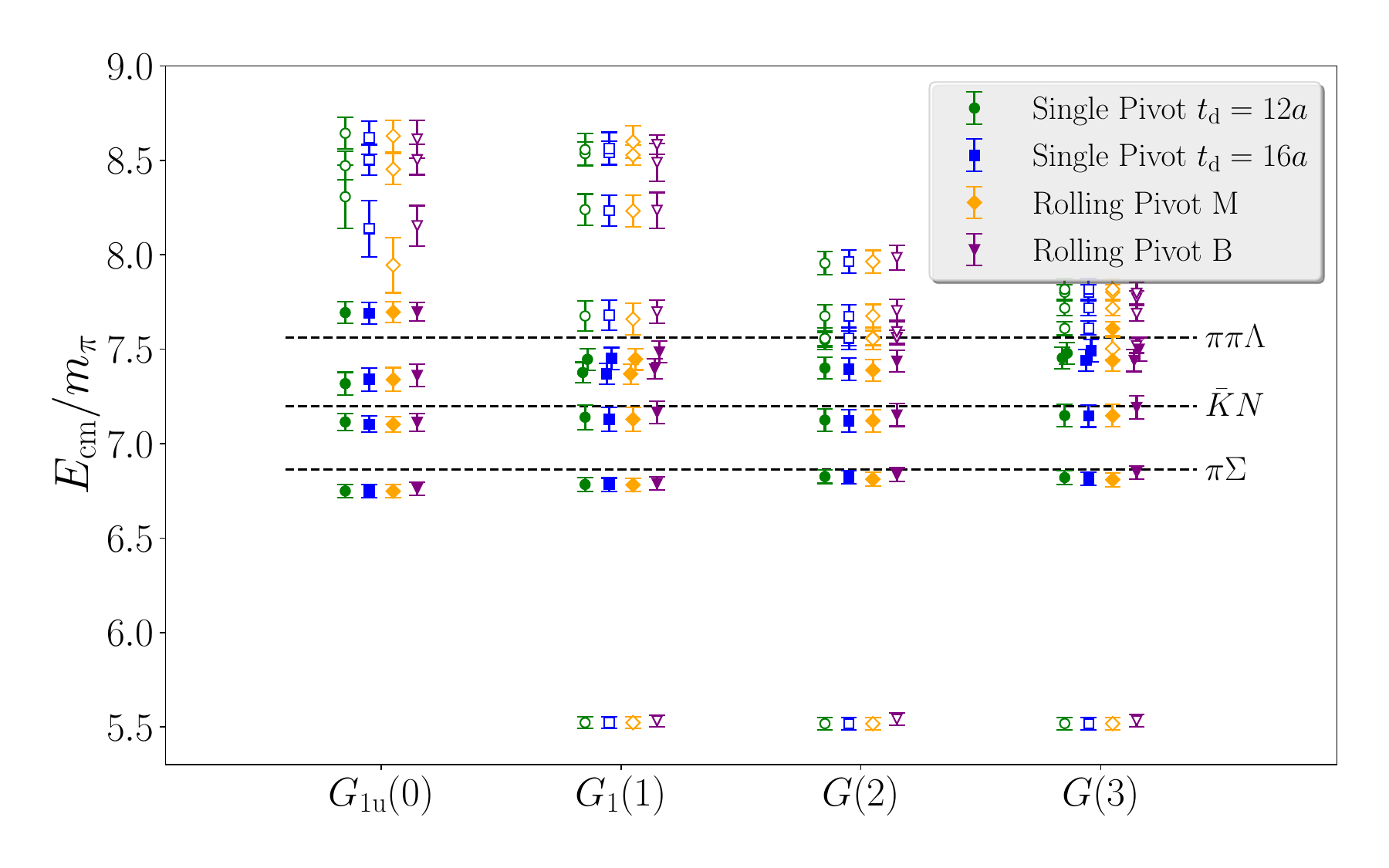}
    \captionsetup{width=.95\textwidth}
    \caption{Center-of-mass finite-volume energy spectra under variation of diagonalization method (single pivot or rolling pivot) and diagonzalization time (examples of $t_{\rm d} = 12a,\, 16a$) for the single pivot method. For the two cases of rolling pivot: (M) The method was implemented on the mean values of the correlators, and the eigenvectors were used to diagonalize the bootstrap samples; (B) The method was implemented on the central value and on the bootstrap samples.}
    \label{fig:center-of-mass-comparison}
\end{figure}

The spectrum results are obtained using two different independent implementations of the variational method: single pivot and rolling pivot. For the single pivot a single choice of $t_{0}$ and $t_{\rm d}$ is used, where the eigenvectors extracted at $t_{\rm d}$ are used to rotate the correlators $C(t)$ at all times $t$, whilst for the rolling pivot a single choice of $t_{0}$ is used, and the correlator $C(t)$ is diagonalized at all times $t$. Figure~\ref{fig:center-of-mass-comparison} depicts the results from both implementations and a variation of $t_{\rm d}$.

\subsection{Finite-volume energies}
\label{section:FiniteVolumeSpectra}

The correlation functions are fitted using variations of tower of exponentials as fit forms. The energies are then determined from correlated-$\chi^{2}$ fits over different $[t_{\rm min}, t_{\rm max}]$ intervals. Single-hadron and multi-hadron correlation functions are treated differently.

\begin{figure}[h!]
    \centering
    \includegraphics[scale=.32]{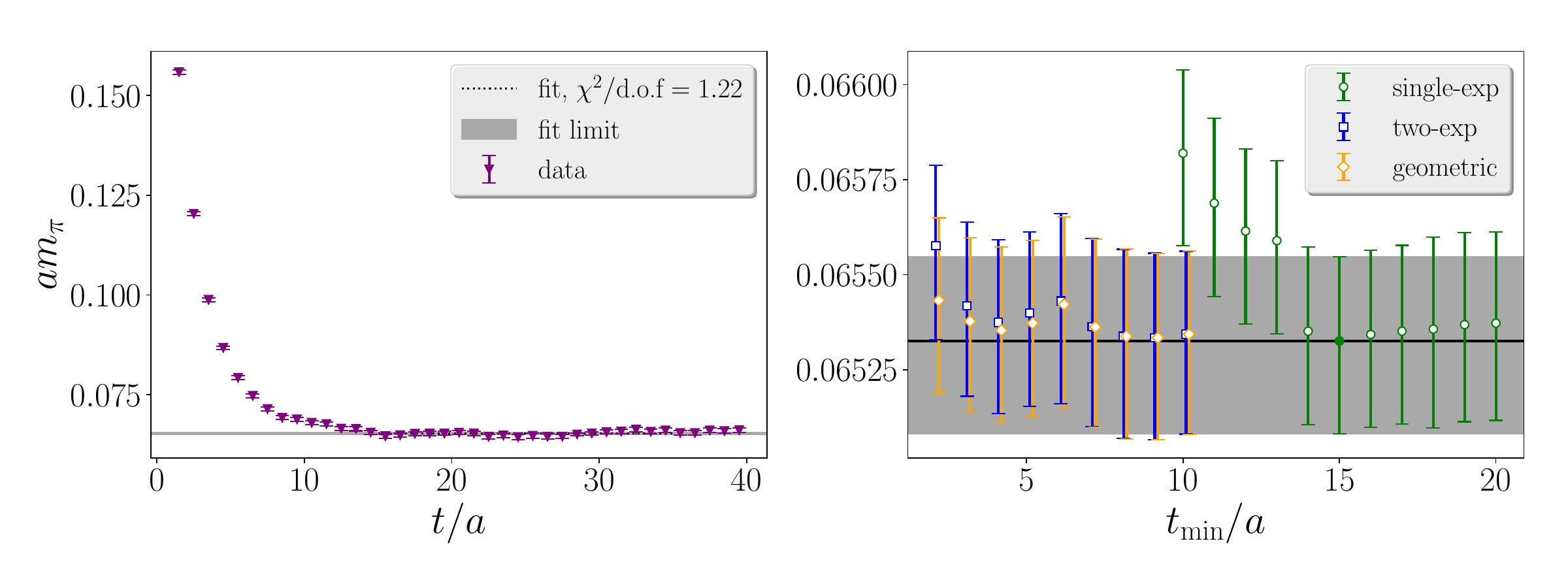}
    \captionsetup{width=.85\textwidth}
    \caption{Pion mass: (left) Effective energy and its final fit result; (right) Different fit models versus variation of $t_{min}$.}
    \label{fig:pion-rest-mass}
\end{figure}

\begin{itemize}
\item[1.] \textit{Single Hadrons:} 
\end{itemize}
\vspace{-.3cm}
Diverse fit models are used, such as one- and two-exponential fits, and geometric exponential series fits. The single-hadron energies correspond to the lowest-lying mesons and baryons. A summary of their masses in lattice units is shown in Table~\ref{tab:hadron-masses}. The chosen $t_{\rm min}$ is based on the consistency of the results from the correlated-$\chi^{2}$ with different fit forms (see Figure~\ref{fig:pion-rest-mass}).

\begin{table}[h]
    \centering
    \begin{tabular}{ll|ll|ll}
    \hline
        $\quad am_{\pi}$  &   $\quad 0.06533(25)\quad$  &   $\quad am_{\Lambda}$  & $\quad 0.3634(14)\quad$ &
        $\quad am_{K}$  &   $\quad 0.15602(16)\quad$   \\ $\quad am_{\Sigma}$  & $\quad 0.3830(19)\quad$ &
        $\quad am_{ N}$  &   $\quad 0.3143(37)\quad$   &  $\quad am_{\Xi}$ &  $\quad 0.41543(96)\quad$ \\
    \hline
    \end{tabular}
    \captionsetup{width=.92\textwidth}
    \caption{Summary of hadron masses in lattice units extracted using exponential fall-offs of the correlation functions of single-hadron operators.}
    \label{tab:hadron-masses}
\end{table}

\begin{figure}[t!]
    \centering
    \includegraphics[scale=.35]{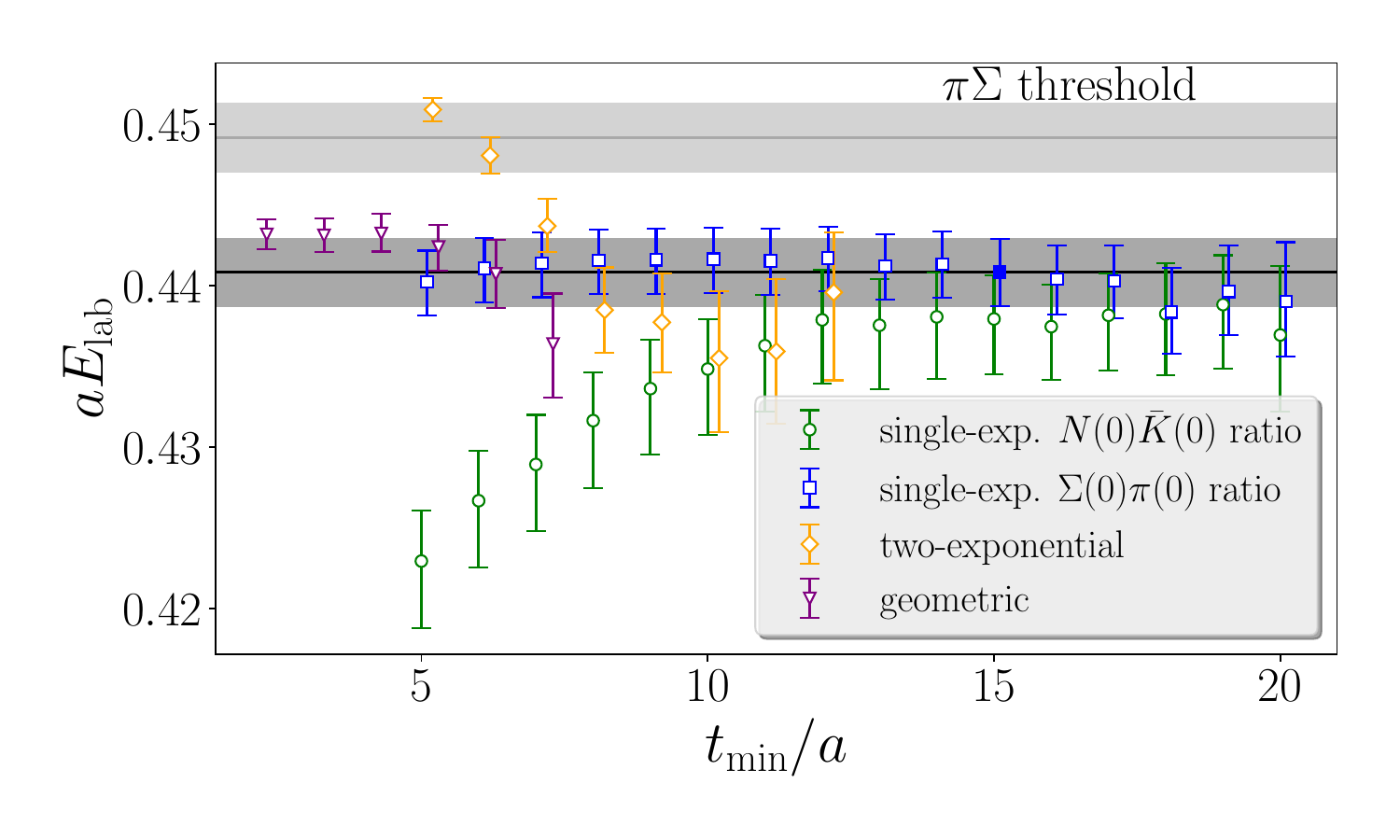}
    \captionsetup{width=\textwidth}
    \caption{Example result: Stability plot of energy fit versus different $t_{\rm min}$ for the lowest level of the $G_{1u}$ irrep using diverse fit models, including two ratios of non-interacting levels ($N(0)\bar{K}(0)$ and $\pi(0)\Sigma(0)$). }
    \label{fig:ratio-correlators}
\end{figure}

\newpage
\begin{itemize}
    \item[2.] \textit{GEVP Eigenvalues:} 
\end{itemize}
\vspace{-.3cm}
Additionally to the fits used for the single-hadrons, one-exponential fits to a ratio of correlators are included for the eigenvalues obtained from the GEVP procedure (see Figure~\ref{fig:ratio-correlators}). This ratio is defined as
\begin{equation}
    R_{n}(t) = \frac{D_{n}(t)}{C_{A}(\mathbf{d}_{A}^{2},t) C_{B}(\mathbf{d}_{B}^{2},t)},
\end{equation}
where $D_{n}(t)$ corresponds to the diagonal of the resulting rotated correlation matrices in the single pivot case, and to the eigenvalues $\lambda_{n}(t)$ in the rolling pivot case; $C_{A,B}$ are the correlation functions of single hadrons, $(A,B) = (\pi, \Sigma)\,\text{or}\,(\bar{K}, N)$ are the thresholds of interest; and $\mathbf{d}_{A,B}^{2}$ are the units of momentum squared for that hadron. This ratio enables the determination of the energy interaction shift $a\Delta E$ from the non-interacting energy $E_{n}^{\rm non-int}$, whilst taking advantage of partial cancellation in the systematic uncertainties.  
\begin{equation}
    E_{n}^{\rm non-int} = \sqrt{ m_{A}^{2} + \left(\frac{2\pi \mathbf{d}_{A}^{2}}{L}\right)^{2} } + \sqrt{  m_{B}^{2} + \left(\frac{2\pi \mathbf{d}_{B}^{2}}{L}\right)^{2}},
\end{equation} 
where $E_{n}^{\rm non-int}$ is the non-interacting energy sum close to the stationary state energy. From this shift $a\Delta E$ the laboratory frame energy $aE_{n}^{\rm lab}$ can be reconstructed as
\begin{equation}
     aE_{n}^{\rm lab} = a\Delta E + aE_{n}^{\rm non-int}.
\end{equation}

\begin{figure}[t!]
    \centering
    \includegraphics[scale=.3]{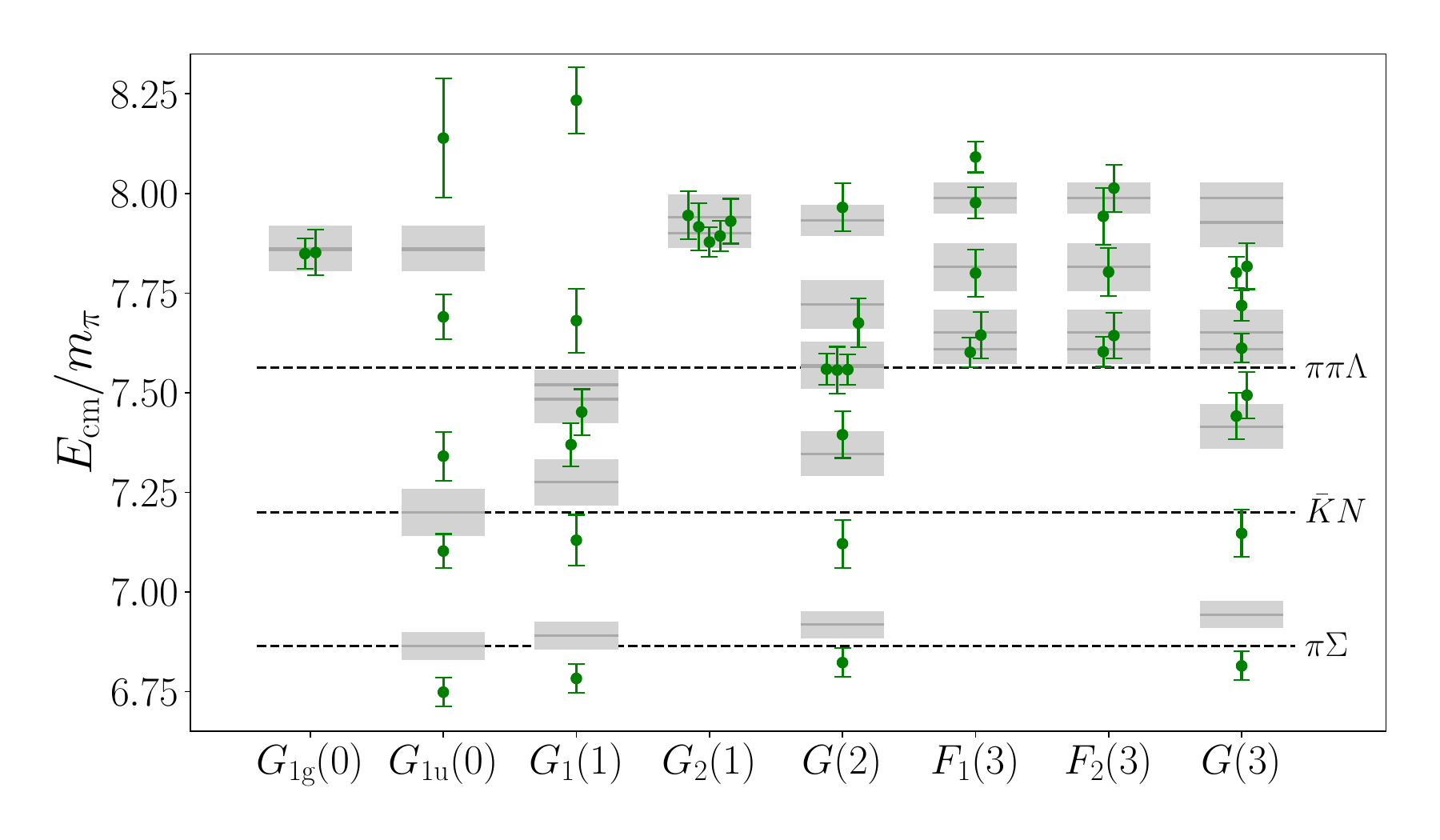}
    \caption{Final results of finite-volume energy spectra, where dashed lines correspond to 
    scattering thresholds; green points are the finite-volume energy spectrum in the center-of-mass frame and their bootstrap error; gray blocks are the locations of energy sums for non-interacting hadrons with increasing momenta. }
    \label{fig:final-results}
\end{figure}

Given that more than one non-interacting energy can be near a certain energy level, and that the reconstructed lab-energy frame should be independent of the choice, all the types of non-interacting levels are used to check consistency of the fit results (see Figure~\ref{fig:ratio-correlators} for an example).

The final spectrum results consist of the finite-volume stationary-state energy spectra shown in Figure ~\ref{fig:final-results}, where all the energy levels extracted from all different irreps are summarized. Results from the relevant irreps are the fundamental physics input that constrains the $\pi\Sigma-\bar{K}N$ coupled-channel scattering-amplitude analysis when employing the L{\"u}scher formalism \cite{Luscher:1990ux,Luscher:1991cf} to explore the $\Lambda(1405)$ energy region.

\section{Conclusion}
\label{section:Conclusion}

This is the first lattice QCD calculation of the $\pi\Sigma-\bar{K}N$ coupled-channel scattering-amplitude. This study was performed using a single ensemble of gauge configurations with $m_{\pi}\approx 200$ MeV and  $a=0.065$ fm. The spectra were reliably extracted using different methods, which consist of variations of the implementation of the GEVP and a variety of fit models, including ratios-of-correlators for diagonalized correlation functions. The spectrum results showed good agreement with the different implementations of the GEVP and consistency with respect to different fit forms. These energy spectra are the key input for the scattering-amplitude analysis via the L{\"u}scher method, which was successfully employed based on the finite-volume stationary-state energy spectra obtained from lattice QCD data, here the final results favored a two-pole picture in the $\Lambda(1405)$ energy region. More details can be found in Refs.~\cite{Bulava:2023gfx, Bulava:2023rmn}.

\subsection*{Acknowledgements}
{\small We thank our colleagues within the CLS consortium for sharing ensembles. Computations were carried out on Frontera at TACC, and at the (NERSC), a U.S. Department of Energy Office of Science User Facility, located at Lawrence Berkeley National Laboratory, 
Contract No. DE-AC02-05CH11231 using NERSC, awards NP-ERCAP0005287, NP-RCAP0010836 and NP-ERCAP0015497. Supported in part by: the U.S. National Science Foundation under awards PHY-1913158 and PHY-2209167 (CJM and SK), the Faculty Early Career Development Program (CAREER) under award PHY-2047185 (AN) and by the Graduate Research Fellowship Program under Grant No. DGE-2040435 (JM); the U.S. Department of Energy, Office of Science, Office of Nuclear Physics, under grant contract numbers DE-SC0011090 and DE-SC0021006 (FRL), DE-SC0012704 (ADH), DE-AC02-05CH11231 (AWL) and within the framework of Scientific Discovery through Advanced Computing (SciDAC) award “Fundamental Nuclear Physics at the Exascale and Beyond” (ADH); the Mauricio and Carlota Botton Fellowship (FRL); and the Heisenberg Programme of the Deutsche Forschungsgemeinschaft project number 454605793 (DM). NumPy \cite{Harris:2020xlr}, matplotlib \cite{Hunter:2007ouj}, and the CHROMA software suite \cite{Edwards:2004sx} were used for analysis, plotting, and correlator evaluation.
}

\end{document}